# Recent Advances and Future Directions in Extended Reality (XR): Exploring AI-Powered Spatial Intelligence

Baichuan Zeng[1,*]

[1]*Department of Computer Science and Engineering, The Chinese University of Hong Kong, Hong Kong*
*\*baichuan@link.cuhk.edu.hk*

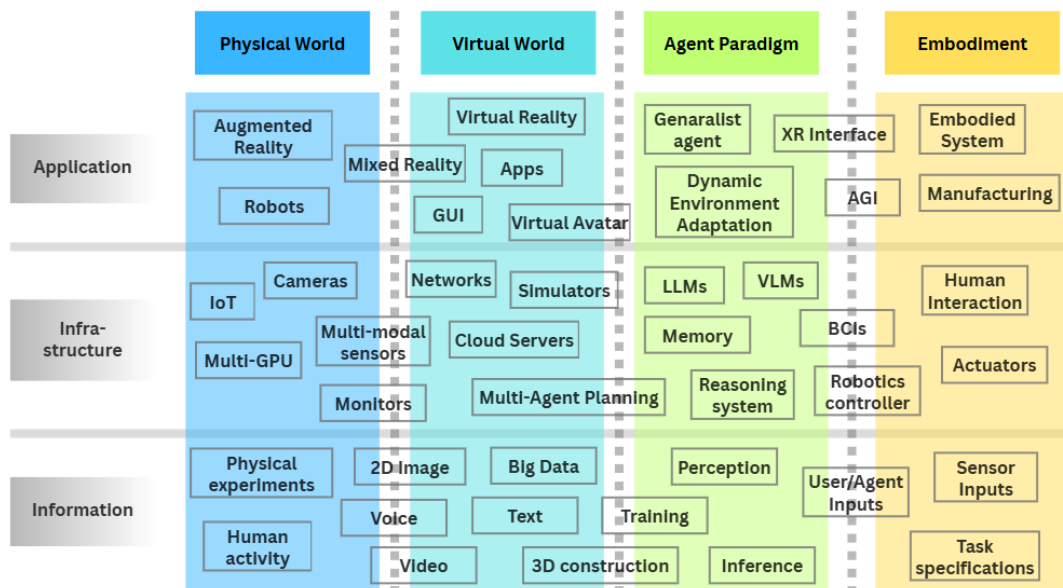

**FIGURE 1.** Overview of XR technology framework with different domains. XR is an exciting way for humans to experience the charm of the digitalized world as well as spatial intelligence. There are many potential agents for XR to integrate to make XR natural and mature, thereby reducing the difference in perception of real and artificial world.

**Abstract.** Extended Reality (XR), encompassing Augmented Reality (AR), Virtual Reality (VR) and Mixed Reality (MR), is a transformative technology bridging the physical and virtual world and it has diverse potential which will be ubiquitous in the future. This review dives into XR's evolution through foundational framework – hardware ranging from monitors to sensors and software ranging from visual tasks to user interface; highlights state of the art (SOTA) XR products with the comparison and analysis of performance base on their foundational framework; elaborates on how commercial XR devices can support the demand of high-quality performance focusing on spatial intelligence. For future expectations, people should pay attention to the integration of multi-modal AI and IoT-driven digital twins to enable adaptive XR system. With the concept of spatial intelligence, XR in future should establish a brand-new space in digits with realistic experience and benefit humanity and human beings. This review underscores the pivotal role of AI in unlocking XR as the next frontier in human-computer interaction.

# INTRODUCTION

Extended Reality (XR) represents a convergence of immersive technology that blends the physical and virtual worlds to varying degrees. Nowadays, there are more and more XR commercial products coming out and this phenomenon shows not only the distance between people and digital world are reduced, but also XR technology will become a trend to assimilate in future life. Augmented Reality (AR), Virtual Reality (VR) and Mixed Reality (MR) are more common terms representing XR, while AR combinates the real world with digital information to assist or augment human perception on the physical world [1], VR provides users with an artificially-constructed environment which is totally immersive [2] and MR is mixed format of AR and VR which provide switching between reality and virtuality. There is also a special term named Augmented Virtuality (AV) which is similar to MR [3] and has the definition literally.

The XR ecosystem can be conceptualized as comprising four primary layers that work together to create immersive experiences. Architecture: The hardware infrastructure that enables XR experiences, including wearable devices, sensors, displays, and the networking components that connect them. Algorithms: The software systems that process sensory input, perform computer vision tasks, and generate appropriate outputs to create convincing XR experiences. User Interface and Experience (UI/UX): The middleware components to perform the world provided by devices, including environmental simulations, rendering systems, and multimodal interaction methods.

The field of XR has experienced remarkable growth, driven by technological advances across multiple domains which includes computer vision, artificial intelligence, sensor technology, and display systems. Some state-of-the-art (SOTA) XR products like Apple Vision Pro (AVP) and Meta Quest 3 (Quest 3) demonstrate the successful integration of these advances. However, despite significant progress, XR still faces considerable challenges, including limited battery life, computational demands for real-time rendering, ergonomic concerns, and issues related to user experience such as cybersickness and limited field of view [4,5]. These challenges present opportunities for innovation and improvement.

This paper examines the current state of XR technology, its applications across various domains, and explores future directions, with a particular focus on how AI-powered spatial intelligence might address existing limitations and enable new capabilities. By analyzing the framework components and showcasing commercial products, this paper aims to provide a comprehensive overview of where XR stands today and where it might be headed in the future.

# TECHNICAL FRAMEWORK

There are three comprehensive layers including Hardware Architecture, Visual Algorithm and UI/UX for XR technology which illustrate through baselines and principles.

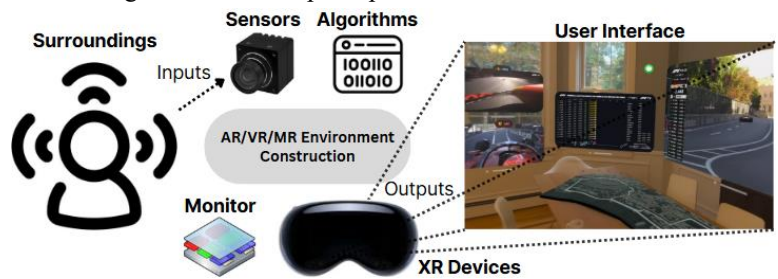

**Figure 2.** Simple illustration of technical framework of XR. For any XR, they are based on physical environment and according to sensorable information they perform in different forms of interfaces.

## Hardware Architecture

The architecture layer of XR systems encompasses the hardware components that form the foundation for immersive experiences. These components must work in concert to capture information about the user and their environment, process this information, and present digital content in a spatially coherent manner.

Display systems in XR must balance resolution, field of view, refresh rate, and form factor. Current technologies include Waveguide Displays in AR which display project images onto transparent optical elements that guide light to the user's eyes while allowing them to see through to the real world [6]. Micro-OLED and Micro-LED Displays offer high pixel density and brightness in compact form factors and are mostly used in SOTA MR products such as AVP [7]. LCD and OLED Panels are common in VR headsets by providing full-field displays with high resolution and refresh rates. For sensor systems, advanced XR devices incorporate multiple sensors to track the user and map the environment. Outward cameras, typically RGB cameras, capture visual information about the environment, and inward cameras or sensors that detect hand positions and gestures. Technologies of depth sensors such as structured light, time-of-flight (ToF), and LiDAR provide spatial mapping capabilities by measuring distances to physical objects. Inertial Measurement Units (IMUs) help track the orientation and movement of the device. In addition, computing architecture is an essential part for XR hardware as for balancing performance, power consumption, and form factors. Standalone devices can incorporate mobile processors to handle on-device rendering, tracking and application logic locally. Split devices distribute computing tasks to external processors instead of headset itself to reduce the load of tasks and improve computing efficiency. Cloud-based processing is an architectures leverage cloud computing for computationally intensive tasks, though this approach must contend with latency challenges [8]. Therefore, higher requirements on connectivity and networking components are necessary. Edge computing serves as an emerging solution and a suitable trade-off for low-latency and strong computing.

The architectural components of XR systems continue to evolve, with miniaturization, power efficiency, and integration, driving progress toward more comfortable and capable devices. Future advances may include specialized XR processors, improved display technologies with higher pixel densities and wider fields of view, and more sophisticated sensor fusion approaches for better environmental understanding.

## Visual Algorithm

The algorithmic layer of XR systems processes sensory input, interprets the user's environment, and generates appropriate outputs to create convincing immersive experiences. These algorithms form the computational backbone of XR, enabling spatial understanding, object recognition, and interactive capabilities.

Computer vision tasks are fundamental to XR, allowing systems to understand and augment the visual environment. Low-level tasks such as Detection, Recognition, Tracking, Segmentation, Simultaneous Localization and Mapping (SLAM), 3D Reconstruction is crucial for manipulating surrounding environments and for further processing. There are already many mature algorithms for achieving these tasks by giving visual input only. For models that require two or more forms of input, such as XR-VIO [9] uses Visual Inertial Odometry (VIO) to achieve faster initialization for XR applications, they demand larger sensory inputs but also have better performance comparing to traditional visual methods. This shows that stronger algorithms cannot leave without the support from hardware architecture. High-level tasks such as gesture recognition, gaze estimation, predictive tracking and AI-accelerated rendering are for natural interactions and immersion. Although some tasks achieved fair performance in XR devices, other tasks like real-time rendering and precise alignment in dynamic environments are still facing issues, which either have an unreal and uncomfortable scenario or heavy equipment.

The algorithmic challenges in XR are significant, particularly in balancing accuracy with real-time performance constraints. Recent advances in hardware-accelerated AI, efficient neural network architectures, and algorithm-hardware co-design are addressing these challenges, enabling more sophisticated XR experiences without sacrificing performance or battery life.

## UI/UX

The user interface layer serves as the middleware between the physical and virtual worlds, providing mechanisms for users to interact with digital content in intuitive ways. Effective XR interfaces must balance usability, learnability, and the unique affordances of spatial computing [10].

Crafting a convincing XR experience hinges on several interconnected components. First, precise calibration and registration are paramount, encompassing spatial alignment, accurate color reproduction, distortion correction, and dynamic registration to seamlessly merge the virtual and real. Algorithms form the bedrock to achieve this precision. Second, environmental simulation is crucial for creating believable virtual worlds, achieved through physically based rendering (PBR), global illumination, spatial audio, and environmental understanding. Third, physical-based interaction engines enhance believability and intuitive interaction by simulating the realistic behavior of virtual objects using rigid body physics and soft body dynamics. Finally, XR interfaces are evolving to incorporate multiple input and output modalities, such as controllers, gestural interfaces, gaze-based interaction, voice commands, haptic feedback, and even brain-computer interfaces (BCIs) [11], to create natural and immersive interaction paradigms.

The UI/UX layer of XR systems continues to evolve as designers explore new interaction paradigms suited to spatial computing. Moving beyond metaphors borrowed from 2D computing toward truly spatial interfaces represents both a challenge and an opportunity for creating more intuitive and powerful XR experiences.

## CASE STUDY AND APPLICATIONS

In this part, SOTA XR products will be introduced with analysis and comparison through the view of aforementioned three levels of technical frameworks.

The XR market has seen significant evolution in recent years, with major technological companies releasing increasingly sophisticated devices. These products represent the SOTA in XR technology and provide insight into different approaches to solving the fundamental challenges of immersive computing. Here is a peek at commercial XR products in Table 1.

**TABLE 1.** Comparative Analysis of Leading XR Headsets [16]

| Feature | AVP [7] | Quest 3 [12] | Varjo XR-4 [13] | Microsoft HoloLens 2 | Xreal Air 2 [14] | Lenovo ThinkReality A3 [15] |
|---|---|---|---|---|---|---|
| Form Factor | Standalone headset | Standalone headset | Tethered headset | Standalone headset | Smart glasses | Tethered smart glasses |
| Display Technology | Micro-OLED | LCD | Mini-LED and Micro-OLED | Waveguide | Micro-OLED | Waveguide |
| Resolution | 23M pixels | 2064×2208 per eye | 3840×3744 per eye | 2048×1080 per eye | 1920×1080 per eye | 1920×1080 per eye |
| Field of View | ~100° diagonal | 110° diagonal | 120° diagonal | 52° diagonal | 46° diagonal | 45° diagonal |
| Tracking | Inside-out 6DoF with LiDAR | Inside-out 6DoF | Inside-out 6DoF | Inside-out 6DoF | 3DoF | 6DoF with host device |
| Hand Tracking | Advanced | Hand tracking | Hand tracking | Hand tracking | Limited | Via host device |
| Eye Tracking | 4 dedicated cameras | None | High precision eye tracking | Basic eye tracking | None | None |
| Processor | Apple M2 + R1 | Snapdragon XR2 Gen 2 | Requires PC | Holographic Processing Unit | None (display only) | Requires PC or smartphone |
| Weight | 650g | 515g | 665g | 566g | 72g | 130g |
| Battery Life | 2-2.5 hours | 2-3 hours | N/A (tethered) | 2-3 hours | Depends on host device | Depends on host device |
| Price (USD) | $3,499 | $499 | $5,990 | $3,500 | $359 | $1,499 |
| Primary Use Case | MR computing | Consumer VR/MR | Professional VR/XR | Enterprise AR | Personal viewing AR | Enterprise AR |

Throughout Table 1, there are some technical highlights to XR products, especially AVP representing the SOTA products with best performance for same tasks among different devices [17][18]. According to a study between Apple and Meta XR products [17], AVP outperformed Quest 3, reducing relative pose error (RPE) and absolute pose error (APE) by 33.9% and 14.6% under various user movements and environmental conditions, respectively. According to a study of comparison of AVP, Quest 3 and Varjo XR-3 [18], for indicators TLX which evaluate cognitive load, CSQ-VR which represents Cybersickness index and Pass-through Quality Metrics, the scores of AVP are better than that of Quest 3, with least performance for Varjo XR-3. The result of comparisons is shown in Figure 4, which APE, RPE, CSQ-VR and TLX are regularized to APE_r, RPE_r, CSQ-VR_r and TLX_r respectively. The reasons causing this solution are significant, one of the most significant facts is that AVP featuring advanced display technology and a sophisticated sensor array. According to Table 1 and Figure 3, AVP assembles: dual micro-OLED displays provide approximately 23 million pixels with a 7.5-micron pixel pitch, creating high visual fidelity; one stereoscopic 3D main camera system, six world-facing tracking cameras, four eye-tracking cameras, and a LiDAR Scanner for precise environmental mapping; dual-chip design with the M2 for general computing and the R1 chip for processing sensor data with minimal latency. Moreover, AVP relies primarily on eye tracking, hand gestures, and voice control rather than physical controllers [17].

However, there are many tread-offs between performance and other components such as weight, price and endurance. From Table 1, although AVP has better performance, it has second highest price (Microsoft HoloLens 2 is out of manufacturing) and is second heaviest among being-sold XR products. For standalone XR devices, their battery life is limited to a range of 3 hours. These showcase that XR devices are still facing fundamental limitations in portability, endurance and comfortability, they are hints for the future direction of development and comparing to generation in the past, current generation already shows significant advances.

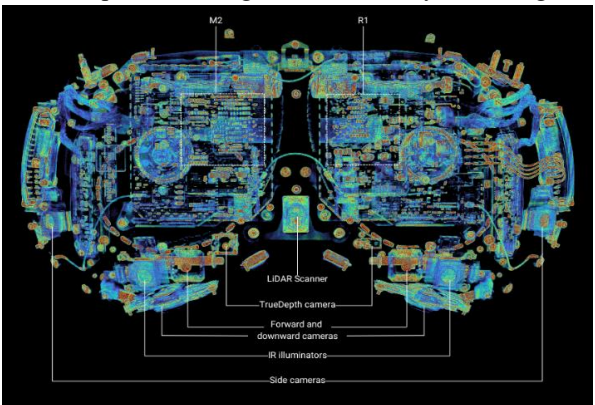

**FIGURE 3.** Front view of internal architecture of AVP which showcases its advantages in sensor system.

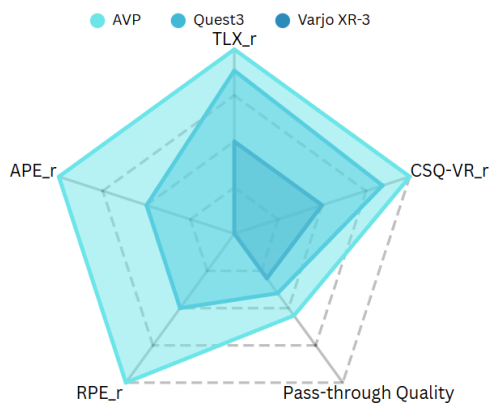

**FIGURE 4.** Comparison among three different XR devices which shows their relative performance. (Varjo XR-3 does not participate in the comparison of APE and RPE)

## DISCUSSION

The potential of XR technology is enormous, this part illustrates the use of XR in daily life and inspires the role of XR in future with respect to the map in Figure 1.

### Applications

XR technologies are finding applications across a wide range of industries, demonstrating versatility and potential for transforming how we work, learn, and interact with digital content. In industry, organizations like Hong Kong Productivity Council (HKPC) leverage XR for workforce development [20], using immersive training simulations to recreate dangerous or expensive scenarios for safety and reduce the cost. In healthcare, medical XR platforms like SurgicalAR by Medivis are transforming healthcare procedures by surgical navigation, anatomical visualization and procedural planning, at the same time the requirement on accuracy on XR is more significant. XR can also be used for entertainment development, architecture design, education and so on. The diversity of XR applications across industries highlights the technology's versatility and suggests its potential for continued expansion into new domains. As technology matures and becomes more accessible, we can expect to see further innovation in how XR is applied to solve real-world problems and create new opportunities for human-computer interaction.

### Spatial Intelligence

It is time to unveil the era of spatial intelligence with the rise of popularity of AI. Delving into the burgeoning domain of spatial intelligence, we observe a convergence of artificial intelligence (AI) and extended reality (XR) technologies that transcend mere immersive experiences. This paradigm shifts empowers systems with the capacity to interpret and adapt to both the surrounding environment and nuanced user needs [21]. Spatial intelligence, therefore, encompasses not only machine perception of the three-dimensional world but also sophisticated interaction and learning within it [22].

The transition from basic visual recognition to a comprehensive understanding and manipulation of the physical realm constitutes a primary objective of spatial intelligence. The evolution of large language models (LLMs) toward multimodal comprehension holds significant ramifications for the XR landscape. Natural language spatial interfaces, for instance, facilitate user control of virtual environment elements through conversational commands and spatial references. This necessitates a profound understanding of physical space to accurately interpret user intent and execute corresponding actions.

Organizations such as World Labs [23] are pioneering the development of AI systems capable of reasoning about spatial relationships and physical interactions in alignment with human intuition. This entails AI's ability to discern not only objects but also the spatial relationships and physical attributes that interconnect them. In smart home contexts, AI assistants can guide users in tasks such as furniture placement or problem-solving, leveraging an understanding of spatial layouts. The concept of digital twins is also undergoing a transformation from static three-dimensional models to dynamic, intelligent representations of physical systems. The integration of Internet of Things (IoT) sensors and XR technologies has yielded spatial perception systems equipped with real-time environmental awareness. Advanced mapping capabilities facilitate persistent and shared spatial understanding, encompassing semantic scene interpretation, dynamic environment tracking, and spatial persistence across devices. Hierarchical spatial memory systems organize spatial information across scales, from granular object details to building layouts and broader geographic areas [21]. These advancements enable XR applications to maintain consistent states and contexts across sessions and devices, fostering a more seamless integration of virtual content with the physical world.

Furthermore, AI systems within XR are exhibiting increasing sensitivity to human behavior and needs. This includes the customization of experiences through the analysis of user preferences and habits, as well as the identification of emotional states via affective computing to dynamically adjust content. In essence, spatial intelligence represents a pivotal frontier in technological advancement [22]. As these technologies mature, XR systems are poised to evolve from passive display mechanisms to active participants in our comprehension of the world around us. This is how XR towards spatial intelligence, by utilizing multi-modal LLM to realize everything spontaneously, instantaneously and vividly.

## CONCLUSION

XR is on the cusp of transforming human-computer interaction, thanks to advances in hardware, algorithms, and user interface design. SOTA products like AVP demonstrate the potential of XR, enabling superior performance in spatial understanding and user experience through advanced display technologies, sensor arrays, and processing. However, challenges remain in balancing performance with factors such as weight, price, battery life and user comfort, highlighting areas where continued innovation is needed.

XR's applications span industries from workforce development and healthcare to entertainment and design, demonstrating their versatility in solving real-world problems and creating new opportunities for immersive experiences. The integration of artificial spatial intelligence represents a significant leap forward, enabling XR systems to not only sense the physical environment, but also understand and interact with it in a way that resembles human intuition.

As XR technology continues to mature, future directions will likely focus on multimodal AI integration, IoT-driven digital twins, and adaptive systems that respond to user behavior and needs. Overcoming existing limitations in portability, durability, and comfort is critical for widespread adoption. Ultimately, XR will evolve from a passive display technology to an integral, active component of our daily lives, changing the way we work, learn, and interact with the world around us.